%% file: base_arxiv.tex
\documentclass[11pt]{article}

\usepackage{graphicx} 
\usepackage{amsmath}
\usepackage{amsthm}
\usepackage{amsfonts}
\usepackage{bm}
\usepackage{amssymb}
\usepackage{xspace}
\usepackage{subfigure}
\usepackage{lineno}
\usepackage{tikz}
\usepackage[]{natbib}
\bibpunct{(}{)}{,}{a}{}{}

\usepackage{multirow}

\usepackage{epic} 
\usepackage{ecltree} 
\usepackage{pstricks} 
\usepackage{pst-node} 
\usepackage{psfrag} 

\usepackage{hyperref}

\usepackage{setspace}
\newcommand{\Mh}{M$_{\text{h}}$\xspace}       

\begin{document}
\title{Hierarchical Modeling of Abundance in Closed Population Capture-Recapture
Models Under Heterogeneity}
\author{Matthew R. Schofield$^{1}$\thanks{Current Address: Department of Statistics, University of Kentucky, Lexington, KY, USA. E-mail: \texttt{matthew.schofield@uky.edu}}~ and Richard J. Barker$^{2}$ \\
\normalsize{$^{1}$Department of Statistics, Columbia University, New York, NY, USA.}\\
\normalsize{$^{2}$Department of Mathematics and Statistics, University of Otago,} \\
\normalsize{P.O. Box 56, Dunedin, New Zealand.}}
\date{}
\maketitle

\begin{abstract}
Hierarchical modeling of abundance in space or time using closed-population
mark-recapture under heterogeneity (model \Mh) presents two challenges: (i)
finding a flexible likelihood in which abundance appears as an explicit
parameter and (ii) fitting the hierarchical model for abundance.  The first
challenge arises because abundance not only indexes the population size, it also
determines the dimension of the capture probabilities in heterogeneity models.
A common approach is to use data augmentation to include these capture probabilities directly into the likelihood and fit the model using Bayesian inference via Markov chain Monte Carlo (MCMC).  Two such examples of this approach are (i) explicit trans-dimensional MCMC, and
(ii) superpopulation data augmentation.  The superpopulation approach has the
advantage of simple specification that is easily implemented in BUGS and related
software. However, it reparameterizes the model so that abundance is no longer
included, except as a derived quantity.  This is a drawback when hierarchical
models for abundance, or related parameters, are desired.  Here, we analytically
compare the two approaches and show that 
they are more closely related than might appear superficially.  We exploit this relationship to specify the model in a way that allows us to include abundance as a parameter and that facilitates hierarchical modeling using readily available software such as BUGS. We use this approach to model trends in grizzly bear abundance in Yellowstone National Park from 1986-1998. 
\end{abstract}

\section{Introduction}
\label{sect:introduction}
Despite their long history and widespread use in ecology and epidemiology
capture-recapture models continue to be the focus of research.  
Usually we treat the closed population problem as being one of inference
about abundance, $N$, a quantity that indexes the size of the population.  While often the
case that a single estimate of $N$ may represent the endpoint of the analysis it is inevitable that analysts will think in terms of describing a collection of abundance estimates across space or time (indexed by $k$).  We use $N$ generically for abundance, both for a single instance or when describing a collection across space or time.  We also use the notation $N_k$ when we wish to be clear that we are referring to members of a larger collection.

One approach to modeling $N$ is the open population model in which a
single population is marked and an integrated model used to describe the
dynamics of the population in terms of recruitment and survival probabilities. 
Here we consider an alternative where closed-population capture-recapture
surveys are replicated in space or time with marked cohorts that are distinct
and hierarchical models are used to describe dynamics of $N$.

\subsection{A motivating example} \cite{Boyce2001} estimated the abundance of
the female grizzly bears, \textit{Ursus arctos}, in the Yellowstone ecosystem
each year in a $13$ year study from $1986$--$1998$ using daily counts of
individually identified bears. To account for sighting heterogeneity (i.e.,
individual variation in detection probabilities), they modeled the capture
frequencies as over-dispersed Poisson counts by fitting a negative binomial
model conditioned on the fact that that the counts are non-zero.  An alternative
approach, which we follow below, is to treat the capture frequencies as summary
statistics from a closed population capture recapture study with $t$ the total
number of days surveyed. 

\cite{Boyce2001} used simulated rank-correlations to
assess the strength of evidence in support of an increasing trend in the numbers
of grizzly bear females.  A more natural approach would be to fit a two-stage
hierarchical model.  In the first stage, we would model female grizzly abundance
each year (indexed by time) using a closed population mark-recapture model
parameterized in terms of $N_k$ and with heterogeneity.  In the second stage we
would model grizzly female abundance as following some sort of trend with
error. Questions concerning long-terms changes in grizzly bear abundance could
then be addressed in terms of parameters of the trend model.

\subsection{Hierarchical capture-recapture models under
heterogeneity}
Hierarchical capture-recapture modeling presents two challenges.  The first is
finding a flexible likelihood in which $N_k$ appears as an explicit
parameter when there is heterogeneity in the capture
probabilities which we denote by $\bm p$. The problem is that the parameters $N$
determine the dimension of $\bm p$. The second challenge is fitting the
hierarchical model for $N$. We focus our initial discussion on this first
challenge but show that it can be resolved in a way that provides a
flexible solution to the second.

The reference model for closed-population capture-recapture under heterogeneity
is \Mh \cite[]{Otis1978}.  A fundamental problem with \Mh is that it is
over-parameterized.  The standard solution \cite[e.g.][]{Burnham1978} is to
model the components of $\bm p$ as random effects drawn from a distribution
indexed by $N$ and to then integrate them from the model.  A drawback however is
that explicit integration of the random effects yields a closed-form likelihood
only in a few restricted cases thereby limiting the number of models
available.  A useful alternative is to integrate over the random effects using
Monte Carlo methods.  One such approach is to model in terms of a complete data
likelihood (CDL) where the likelihood is `completed' by including the unobserved
$\bm p$; inference can then proceed via the the (Monte Carlo) EM algorithm
\cite[]{Dempster1977,Levine2001a} or the Gibbs sampler \cite[]{Geman1984}.  

An impediment to use of the CDL is that its dimension changes with $N$ through
$\bm p$.  Three seemingly different approaches have been used to circumvent this
problem.  
The first is to eliminate the dependence of $N$ on $\bm
p$ in a Gibbs sampler by integrating out $\bm p$ from the joint full conditional
distribution of $N$ and $\bm p$ \cite[]{Fienberg1999}.  Here, we do not consider combining MCMC and other numerical integration techniques, although we briefly discuss the use of quadrature in evaluating the likelihood in section \ref{sec:discussion}.

The second approach is to use a 
trans-dimensional (TD) algorithm to update the parameter $N$.  
\cite{Sisson2005} defines a TD algorithm as one that admits transitions
between states that have corresponding parameter vectors of differing dimension (in our case the states are indexed by values of $N$).  One example of a TD algorithm is the reversible jump Markov chain Monte Carlo (RJMCMC) algorithm \cite[]{Green1995} that explicitly accounts for the differing dimension when moving between states.  Another example is the product space approach of \cite{Chib1995} that avoids the need to explicitly account for the differing dimension when moving between states.  
Such TD algorithms have been used by \cite{Fienberg1999}, \cite{King2008}, \cite{Durban2005} (subsequently DE) and others in the context of \Mh or related capture-recapture models.

The third approach is to use superpopulation data augmentation to specify the model.  The term `superpopulation' refers to an approach in which the likelihood is completed by augmenting $\bm p$ in a way that fixes its dimension.  The phrase `data augmentation' (DA) refers to a family of computational methods that add missing data/auxiliary variables into the likelihood \cite[]{Gelman2004c} to facilitate modeling.   
DA is used, both for the CDL models mentioned above, as well as in the superpopulation models of \cite{Royle2007} (subsequently RDL), \cite{Royle2008}, \cite{Link2010} and others in the context of \Mh or related capture-recapture models.  The idea behind the approach of RDL came from identifying a relationship between capture-recapture and occupancy modeling \cite[]{MacKenzie2002}.  

While elegantly simple, the approach suggested by RDL is unnatural for capture-recapture modeling in that it relegates $N$, the parameter of interest, to the status of derived parameter.  A practical consequence is that in Bayesian modeling it can make prior specification difficult, particularly for hierarchical models in which interest is in modeling $N$ through space and time.  
\subsection{Outline of the rest of the paper}
Our purpose is two-fold.  Firstly, to provide a critical examination of the
methodological issues underlying various augmented data representations of
closed population capture-recapture models when capture probabilities are
heterogeneous (approaches two and three above). In
Section \ref{sect:comparelike} we review integrated likelihood, full-likelihood
and super-population data augmentation and investigate their relationship and
how parameters can be modeled within a Bayesian framework.  Our second purpose
is to show how the CDL can be explicitly expressed in terms of the abundance
parameter $N$ for the capture recapture model \Mh in a manner that is
straight-forward to implement using generic Bayesian hierarchical modeling. In
section
\ref{sec:hierarchical} we introduce our hierarchical extensions to \Mh 
and illustrate these methods using the grizzly bear example and show how to fit
this model using the readily available software BUGS \cite[]{Lunn2000} or JAGS
\cite[]{Plummer2003} for carrying out Bayesian analysis of data.   We offer
concluding comment in Section \ref{sec:discussion}.



\section{\Mh Likelihoods and varying dimensions}\label{sect:comparelike}
Likelihoods for model \Mh include an integrated likelihood \cite[]{Burnham1978}, a ``full'' likelihood \cite[]{King2008}, and a super-population likelihood of RDL.  The three most obvious differences between these approaches are (i) whether or not $N$ is explicitly included in the likelihood, (ii) the presence or absence of combinatorial terms involving $N$, and (iii) the presence or absence of a superpopulation (with corresponding value $M$) in the model.  We do not consider conditional approaches, such as those discussed by \cite{Sanathanan1972}.

In the following we use $[\bm x | \bm p]$ to denote the model for the latent data $\bm x$ given the vector of capture probabilities $\bm p$.  Unless we specify otherwise, $x_{ij}$ is 1 if individual $i$ $(i=1,\ldots N)$ was captured in sample $j$ $(j=1,\ldots,t)$ and 0 otherwise, and $p_{i}$ is the capture probability of individual $i$.  
We also consider the observed data $\bm x^{obs}$, where $x^{obs}_{ij}$ is $1$ if individual $i = 1,\ldots,n$ was captured in sample $j$, where $n$ is the number of distinct individuals caught in the study.  
Other than the inclusion of unobserved individuals in $\bm x$, the only difference between $\bm x$ and $\bm x^{obs}$ is being unable to align the row indices of $\bm x^{obs}$ with the row indices of $\bm x$.  There are $\frac{N!}{\prod_{h}z_{h}!}$ possible $\bm x$ matrices that are compatible with $\bm x^{obs}$ for a given value of $N\geq n$, where $z_h$ is the number of individuals observed with capture history $h$.  Note that $z_{0} = N-n$ denotes the number of individuals with the null history $00\ldots0$.  

\subsection{Burnham's Integrated Likelihood}
Burnham's integrated likelihood \cite[]{Burnham1978} can be considered as the
standard form of the observed data likelihood (ODL) for model \Mh.  The ODL is
obtained by modeling the unknown capture probabilities $\bm p$ as exchangeable
random effects sampled from a distribution with parameters $\theta_{p}$, $[\bm p
| \theta_p]$ with $cdf$ $F(\bm{p}) \equiv F_{\theta_p}(\bm{p})$.  Starting with
\begin{equation}\label{eq:XgivenP}
 [\bm x | \bm p] = \prod_{i=1}^N \prod_{j=1}^{t} p_i^{x_{ij}}(1-p_i)^{1-x_{ij}} 
\end{equation}
and summing over the permutations of $\bm x$ and integrating out $\bm{p}$ leads to:
\begin{equation}
 \label{eq:burnham}
 [\bm{x}^{obs}|\theta_{p},N]  = \frac{N!}{\prod_{h}z_h!} \prod_{j=0}^t \xi_{jF}^{f_j}
\end{equation}
where $f_{j}$ is the number of individuals caught $j$ times, and $\xi_{jF} = \int_0^1 p^j(1-p)^{t-j}dF(\bm p)$.  Our focus here is on parametric distributions $F$ that do not allow $\xi_{jF}$ to be evaluated directly.  \cite{Tardella2002} and \cite{Farcomeni2010} consider a case where $F$ is a finite-dimensional non-parametric distribution and the resulting MCMC algorithm does not require a TD step.

\subsection{Complete Data Likelihood (CDL) - $N$ as a parameter}
Starting with (\ref{eq:XgivenP}) and including a hierarchical model for $\bm p$, we can write a complete data likelihood as
\begin{equation}
 [\bm{x}^{obs},\bm{p}|\theta_{p},N] = \underbrace{\frac{N!}{\prod_{h}z_h!} ~ {
\prod_{i=1}^{n} \prod_{j=1}^{t}
p_i^{x^{obs}_{ij}}(1-p_i)^{1-x^{obs}_{ij}}\prod_{i=n+1}^{N}(1-p_i)^{t}}}_{[\bm{x
}^{obs}|\bm{p},N]} \underbrace{\prod_{i=1}^{N}[p_i | \theta_p]}_{[\bm
p|\theta_{p},N]}  \label{eq:CDL2}
\end{equation}
If we integrate across $p_{i},~i=1,\ldots,N$ we obtain equation (\ref{eq:burnham}). 
We refer to (\ref{eq:CDL2}) as the CDL-$N$ model.  More details of CDL-$N$ are available in the supplementary materials, section 1.1.
\subsection{Complete Data Likelihood (CDL) - $w$ as a parameter}
RDL describe a complete data likelihood in terms of a super-population $M$, 
with $\bm x$ now an $M$ by $t$ matrix.  The idea is that we replace $N$ in the likelihood by 
 $\bm w$, a vector of binary labels 
vector of dimension $M$ where $w_{i}$ takes the value $1$ if individual $i$ is
included in the population and $0$ otherwise such that $N=\sum_{i=1}^{M}w_{i}$. 
 RDL model the elements $w_i$ of $\bm w$ using an exchangeable Bernoulli prior
with parameter $\phi$.  Thus, their complete model can be written as
\begin{align}
 [\bm x, \bm w, \bm p | \theta_p,\phi] &= [\bm w | \phi][\bm p | \bm w, \theta_p][\bm x |\bm w, \bm p]\nonumber\\
 &= [\bm w | \phi][\bm p | \theta_p][\bm x |\bm w, \bm p] \nonumber\\
 &=\prod_{i=1}^M \left\{ [p_i|\theta_{p}] \phi^{w_i}(1-\phi)^{1-w_i} \prod_{j=1}^t (p_iw_i)^{x_{ij}} (1-p_iw_i)^{1-x_{ij}}\right\},\label{eq:spCDL}
\end{align}
where $\phi$ is the probability that an individual in the superpopulation (of size $M$) is also a member of the population (of size $N$).

Two observations on this approach are relevant for subsequent discussion.  The first is that the RDL approach features two distinct data augmentation steps, one involving $\bm w$ and a second involving $\bm p$. The first DA step involving $\bm w$ is equivalent to replacing the scalar parameter $N$ in the usual formulation of model \Mh with the vector parameter $\bm w$.  The second DA step specifies a value $p_{i}$ $(i=1,\ldots,M)$ for all individuals in the superpopulation, whether they are in the population or not.
 
The second observation is that starting with (\ref{eq:spCDL}), then integrating over $\bm{p}$, summing over $\bm w$, and finally summing over permutations of $\bm x$ leads to
\begin{equation}
 \label{eq:andyintegrated}
 [\bm{x}^{obs},N|\theta_{p},\phi,M] = \underbrace{\frac{N!}{\prod_{h}z_h!}{\prod_{j=0}^t \xi_{jF}^{f_j}}}_{[\bm{x}^{obs}|\theta_{p},N]} \times \underbrace{\binom{M}{N}\phi^N(1-\phi)^{M-N}}_{[N|\phi,M]},
\end{equation}
which is equivalent to the ODL in (\ref{eq:burnham}) multiplied by the prior for
$N$ induced by modeling the set of values $w_i$ as conditionally exchangeable
Bernoulli random variables with parameter $\phi$. 
Thus, replacing $N$ with $\bm{w}$ leads to a Bayesian reparameterization of \Mh in which the implied prior on $N$ is specified by a binomial distribution.  We refer to (\ref{eq:spCDL}) as the CDL-$w$ model.  More details of CDL-$w$, including the move from (\ref{eq:spCDL}) to (\ref{eq:andyintegrated}) are in the supplementary materials, sections 1.2 and 1.3.

Note that RDL do not use (\ref{eq:andyintegrated}) to make inference. Instead, they only consider one partitioning of $\bm x$, where $\bm x^{obs}$ is augmented by an $(M-n) \times t$ matrix of zeros.  This leads to an integrated likelihood proportional to that shown in (\ref{eq:andyintegrated}), see supplementary materials, section 1.3 for details.

\subsection{Model Fitting}
Modeling via the ODL in (\ref{eq:burnham}) is straight-forward with the trans-dimensional nature of the model eliminated, provided we can explicitly integrate over the random effects distribution of $\bm p$.  Unfortunately we can only do this for a restricted choice of random effects distribution.  We can solve this problem through use of Bayesian CDL approaches and Markov chain Monte Carlo (MCMC) for model fitting.

The CDL-$w$ model can be easily implemented using Gibbs sampling.  An associated advantage is the simplicity of specifying model in BUGS.  We refer the reader to \cite{Royle2007} for details of the Gibbs sampler for the CDL-$w$ model, including implementation in BUGS.  
A point we will return to later is that it is potentially difficult to include flexible priors for $N$ in the CDL-$w$ model. 
In particular, it is difficult to include any prior that is not a binomial prior that is induced through the specification outlined earlier.  This is unlikely to be a problem when we have a single closed population data set.  However, it becomes more important if we have replicated closed population data sets, across space or time, for which we wish to specify a hierarchical model for $N$.

Here we outline several approaches that can be used to fit model CDL-$N$.  As already noted, the problem is that the random variable $N$ defines the dimension of the parameter vector $\bm p$.  To overcome this \cite{King2008} implement a reversible-jump MCMC algorithm to update $N$ within a MCMC procedure.  All other unknowns can be updated using standard Gibbs sampling steps.  

A different approach was implemented by DE who use the algorithm of \cite{Carlin1995}.  A necessary part of this algorithm is to specify an upper size limit $M$ and values $p_{N+1},\ldots,p_{M}$ in the same way as in the CDL-$w$ model.  To complete specification of the algorithm, DE specify `pseudo-priors' for the values of $p_{N+1},\ldots,p_{M}$.  The pseudo-priors are not expressed in the likelihood and their sole purpose is to improve the efficiency of the algorithm.  DE used pilot runs of the algorithm to determine appropriate pseudo-priors.

The fitting of CDL-$N$ has been critized by RDL due to the need to move between `models' using an explicit TD step -- either (i) the reversible-jump step of \cite{King2008}, or (ii) the need for pseudo-priors (and pilot runs) in DE \cite[]{Royle2007,Royle2012}.  To show this criticism is unjustified, we consider an additional approach to fitting the CDL-$N$ model.  It amounts to a special case of the DE algorithm, where we overcome the need for a pilot run by choosing the pseudo-prior for $p_{i},~i>N$ to be the hierarchical distribution for $p_{i},~i=1,\ldots,N$.  Such an approach appears to share much in common with the approach used to fit CDL-$w$: both include an upper size limit $M$ and specify the same model for `ghost' individuals (i.e. those individuals that are in $M$ but not $N$).   Details of the algorithm as well as associated BUGS code is in the supplementary materials, sections 2.1 and 3.

A closer comparison between the algorithms used to fit CDL-$N$ and CDL-$w$ reveals that 
the problem of moving between models does exist in the CDL-$w$ formulation.  The key to understanding this is to consider model fitting using CDL-$N$ as a problem of selecting between models, each indexed by a different value for the parameter $N$.  Similarly, model fitting using CDL-$w$ is a problem of selecting between models, each indexed by a different value for $w$. 
Moreover, the MCMC algorithm used to fit CDL-$w$ is identical to a specific TD algorithm for a model where we (i) reparameterize model \Mh by replacing the scalar parameter $N$ by the vector parameter $\bm w$, including the capture probabilities $p_{i}$ in the likelihood for individuals included in the population (i.e. with $w_i=1$), and (ii) assume the choice of distribution for augmenting variables has been fixed.  
The distinction between the likelihood for (i) above and CDL-$w$ is subtle; in (i) the capture probabilities $p_{i}$ for the ghost individuals ($w_{i}=0$) are not included in the likelihood as they are not defined.
 More details of the equivalence of the two algorithms, including showing that pseudo-priors are a latent feature in the model of RDL as well as the corresponding MCMC algorithms are in the supplementary materials, sections 2.2 and 4.



\section{Hierarchical Extensions}\label{sec:hierarchical}
We are interested in situations where replicated closed population data sets are observed through time or across space.  
In general, we suppose that $m$ distinct closed-population studies have been
carried out and that we are unable to link individuals from one study to
another.  In addition to estimating abundance from each of these data sets,
we also wish to model changes in abundance. We outline the changes required in
the CDL-$N$ and CDL-$w$ models to include this extension below.

\subsection{CDL-$N$}
The latent capture history matrix for the $k$th data set is denoted by the $N_{k} \times t_{k}$ matrix $\bm x_k,~k=1,\ldots,m$, where $N_{k}$ is the abundance for data set $k$ and $t_{k}$ is the number of samples conducted in data set $k$.  The value $x_{ijk} = 1$ if individual $i$ in sample $j$ of data set $k$ was observed and $x_{ijk}=0$ otherwise.  Note that individual $i=i^{*}$ associated with data set $k=k^{*}$ is not linked to individual $i=i^{*}$ associated with data set $k\neq k^{*}$.  Likewise, the observed capture history matrix for data set $k$ is denoted by the $n_{k} \times t_{k}$ matrix $\bm x^{obs}_{k},~k=1,\ldots,m$.  
In general we model the data from the $m$ data sets as being exchangeable given the parameters,
\begin{align*}
 &[\bm{x}^{obs}_{1},\ldots,\bm x^{obs}_{m},\bm{p}_{1},\ldots,\bm{p}_{m}|\theta_{p},N_{1},\ldots,N_{m}] = \prod_{k=1}^{m}[\bm{x}^{obs}_{k},\bm{p}_{k}|\theta_{p},N_{k}]\\
 &\hspace{0.5cm}= \prod_{k=1}^{m}\Bigg\{\underbrace{\frac{N_k!}{\prod_{h}z_{hk}!} ~ \prod_{i=1}^{n_k} \prod_{j=1}^{t_{k}} p_{ik}^{x_{ijk}}(1-p_{ik})^{1-x_{ijk}}\prod_{i=n_{k}+1}^{N_{k}}(1-p_{ik})^{t}}_{[\bm{x}^{obs}_k|\bm{p}_k,N_k]} \underbrace{\prod_{i=1}^{N}[p_{ik} | \theta_p]}_{[\bm p_k|\theta_{p},N_k]}\Bigg\},
\end{align*}
where $\bm p_k$ are the vector of capture probabilities from data set $k$ and $z_{hk}$ is the number of individuals caught with capture history $h$ in data set $k$.  That is, we model the $m$ distinct data sets separately as described in section \ref{sect:comparelike}, although we may choose to specify hierarchical models on the parameters of these models.

Instead of keeping the model general, we focus on allowing a hierarchical model for $N_k$ that includes a trend for the abundance through time.  One such possible model is a Poisson model with parameter $\lambda_{k}$,
\[
 [N_{k}|\lambda_{k}] = \prod_{k=1}^{m} \frac{\lambda_{k}^{N_{k}}\exp(-\lambda_{k})}{N_{k}!}
\]
with a model for $\log(\lambda_{k})$ that is a linear regression through time,
\[
 [\log(\lambda_{h})|\bm\beta,\sigma_{\lambda}] = \prod_{h=1}^{k}\frac{1}{\sqrt{2\pi}\sigma_{\lambda}}\exp\left(\frac{-1}{2\sigma_{\lambda}^{2}}(\log(\lambda_{h})-\beta_{0}-\beta_{1}h)^2\right).
\]
This model includes $\sigma_{\lambda}$ which accounts for overdispersion in the
Poisson model for $N_{k}$.
Note that this model is useful to illustrate the inclusion of a hierarchical model for $N_{k}$ but we do not consider this the definitive model for modeling abundance through time.  Many other approaches are reasonable including different choices of distribution and trend model.

To fit this hierarchical extension requires a standard extension of the previous MCMC algorithm for model CDL-$N$.  We can use a Gibbs sampler and update those additional parameters with known full conditional distributions directly.  If a parameter cannot be updated by sampling from a known distribution we use an algorithm that allows sampling from unnormalized density (e.g., the Metropolis-Hastings algorithm).  Such extensions are easily included in programs such as BUGS.

\subsection{CDL-$w$}\label{sec:cdlw}
As with CDL-$N$, we extend the capture history matrices to allow for replicated data sets.  The only difference from CDL-$N$ is that the dimension of the latent capture history matrix is $M_{k} \times t_{k}$ for the $k$th data set, where $M_{k}$ is the superpopulation for data set $k$.  We also extend the vector $\bm w$ to allow for multiple data sets, with $\bm w_k$ denoting the vector for the $k$th data set.  The value $w_{ik} = 1$ if individual $i$ in data set $k$ is included in the population and $w_{ik}=0$ otherwise.

As with CDL-$N$ we model $\bm x_{k}$ and $\bm p_{k}$ from the $m$ data sets  as being exchangeable given the parameters,
\begin{align*}
 [\bm x_{1},\ldots,\bm x_{m},\bm p_1,\ldots,\bm p_m|\bm w_{1},\ldots,\bm w_{m},\theta_p,\phi_1,\ldots,\phi_m] = \prod_{k=1}^{m}[\bm x_{k}|,\bm w_{k},\bm p_k][\bm p_{k}|\bm w_{k},\theta_p]&\\
= \prod_{k=1}^{m}\left\{ \prod_{i=1}^{M_{k}} \left\{ [p_{ik}|\theta_{p}] \prod_{j=1}^{t_{k}} (p_{ik}w_{ik})^{x_{ijk}} (1-p_{ik}w_{ik})^{1-x_{ijk}}  \right\}\right\}&
\end{align*}

Including the required hierarchical model for $N_{k}$ is more difficult for CDL-$w$.  
The problem is that we need to express the model for $N_{k}$ in terms of the random variables $\bm w_{k}$.  That is, given a desired model (or prior) on $N_{k}$, with density $[N_{k}|\lambda]$, we seek a distribution $[\bm w_{k}|\lambda]$, such that
\[
 [w_{\cdot k}|\lambda] = [N_{k}|\lambda],
\]
where $w_{\cdot k} = \sum_{i}w_{ik} \equiv N_{k}$.  For this specific case, we need to specify a distribution $[\bm w_{k}|\lambda]$ that induces
\[
 [w_{\cdot k}|\lambda] = [N_{k}|\lambda] = \frac{\lambda_{k}^{N_{k}}\exp(-\lambda_{k})}{N_{k}!}.
\]
One such distribution that satisfies this is
\begin{equation}\label{eq:wdist1}
 [\bm{w}_{k}|\lambda_{k}] \propto \lambda_{k}^{w_{\cdot k}}\exp(-\lambda_{k})\left(M_{k}-w_{\cdot k}\right)!
\end{equation}
since we know that for any particular value of $N_{k}$ there are ${M_{k} \choose N_{k}}$ realizations of $\bm w_{k}$ that satisfy $\sum_{i}w_{ik} = N_{k}$.

An alternative approach is to use parameter expansion \cite[]{Liu1999}; that is, we add a (non-identifiable) parameter to the distribution to facilitate modeling \cite[]{Gelman2004}.  In particular, we want to specify the distribution for the indicator variables $\bm w_{k}$ such that when we integrate over the additional parameter $\phi_{k}$ we obtain
\[
 \int_{\phi_{k}}[w_{\cdot k}|\phi_{k},\lambda][\phi_{k}]d\phi_{k} = [w_{\cdot k}|\lambda] = [N_{k}|\lambda].
\]
One example that satisfies this requirement is
\begin{equation}\label{eq:wdist2}
  [\bm{w}_{k}|\phi_{k},\lambda_{k}] \propto \prod_{i=1}^{M_{k}}\left\{\phi_k^{w_{ik}}(1-\phi_{k})^{1-w_{ik}}\right\}\frac{\lambda_{k}^{w_{\cdot k}}\exp(-\lambda_{k})}{\left(w_{\cdot k}\right)!}
\end{equation}
with $\phi_{k}$ having a uniform distribution between $0$ and $1$.

Once we have specified the distribution for $\bm w_{k}$ to obtain the required model, we include the distribution for $\log(\lambda_{k})$ as in the CDL-$N$ model,
\[
 [\log(\lambda_{h})|\bm\beta,\sigma_{\lambda}] = \prod_{h=1}^{k}\frac{1}{\sqrt{2\pi}\sigma_{\lambda}}\exp\left(\frac{-1}{2\sigma_{\lambda}^{2}}(\log(\lambda_{h})-\beta_{0}-\beta_{1}h)^2\right).
\]

How we include this hierarchical extension in model fitting depends on the method of fitting.  If we are using self-written code, we can fit the model using either (\ref{eq:wdist1}) or (\ref{eq:wdist2}) above.  However, to the best of our knowledge, we are unable to specify the model using (\ref{eq:wdist1}) in BUGS.  We can, however, include (\ref{eq:wdist2}) using a ``ones/zeros trick'' as described in the supplementary materials section 5.

\subsection{Example: Grizzly Bear Abundance} \label{sec:grizzlies}
\cite{Boyce2001} estimated the abundance of the female grizzly bears, \textit{Ursus arctos}, in the Yellowstone ecosystem each year in a $m=13$ year study from $1986$--$1998$ using negative binomial models on the counts of individual bears.  Instead of modeling the counts we treat the counts as arising from $t$ capture-recapture samples in each year (across which we assume closure).  Unfortunately, the number of sampling occasions is not specified (and in fact may be ill-defined), but the methods section of the manuscript suggests that we could treat the sampling as undertaken each day during summer resulting in $t \approx 150$.  To test the sensitivity of the model to the value of $t$ we also fit the model with $t=20$.

We use the hierarchical model specified above with one additional extension.  We model the capture probabilities within each data set as being exchangeable random variables from a logit-normal distribution,
\[
  \text{logit}(p_{ik}) \stackrel{exch.}{\sim} \mathcal{N}(\mu_{k},\sigma^{2}_{p}),~~i=1,\ldots,N_k,~k=1,\ldots,m,
\]
and model the mean values by an exchangeable normal prior,
\[
  \mu_{k} \stackrel{exch.}{\sim}
\mathcal{N}(\mu_{\mu},\sigma^{2}_{\mu}),~~k=1,\ldots,m.
\]

To complete the specification we include prior distributions.  We place normal prior distributions on $\beta_{0}$ and $\beta_{1}$ with mean of $0$ and variance $100000$,  a logistic prior distribution on $\mu_{\mu}$ with location $0$ and scale $1$ and Student's $t$ prior distributions (truncated above zero) on $\sigma_{p}$, $\sigma_{\mu}$ and $\sigma_{\lambda}$ with location $0$, scale $25$ and $3$ degrees of freedom.  The prior distributions for the standard deviations were specified using the redundant parameterization of \cite{Gelman2006}.
We fitted the model in JAGS, with full details of the models in the JAGS code at \url{www.maramatanga.com}.

\begin{figure}[!htbp]

 \makebox[\textwidth]{
 \input{grizzlycomp}
 }
 \caption{Posterior distribution estimates for abundance through time for female grizzly bears in the Yellowstone 
ecosystem between 1986--1998 from fitting model \Mh to the data of Boyce et. al
(2001). For the models with $t=150$ and $t=20$, the thin vertical lines span the central 95\% credible interval, the thick vertical line span the quartiles  of
the posterior distribution and the central symbol corresponds to the median.  For the Boyce model, the thin vertical lines represent the 95\% confidence interval and the central symbol corresponds to the point estimate found in Boyce et. al. (2001).  The plotted trend for $N$ is the median, $2.5\%$ and $97.5\%$ quantiles of
the posterior distribution for $\exp(\beta_{0} + \beta_{1}h)$ for the models with $t=150$ (black) and $t=20$ (gray).  } \label{fig:NGrizz}
\end{figure}
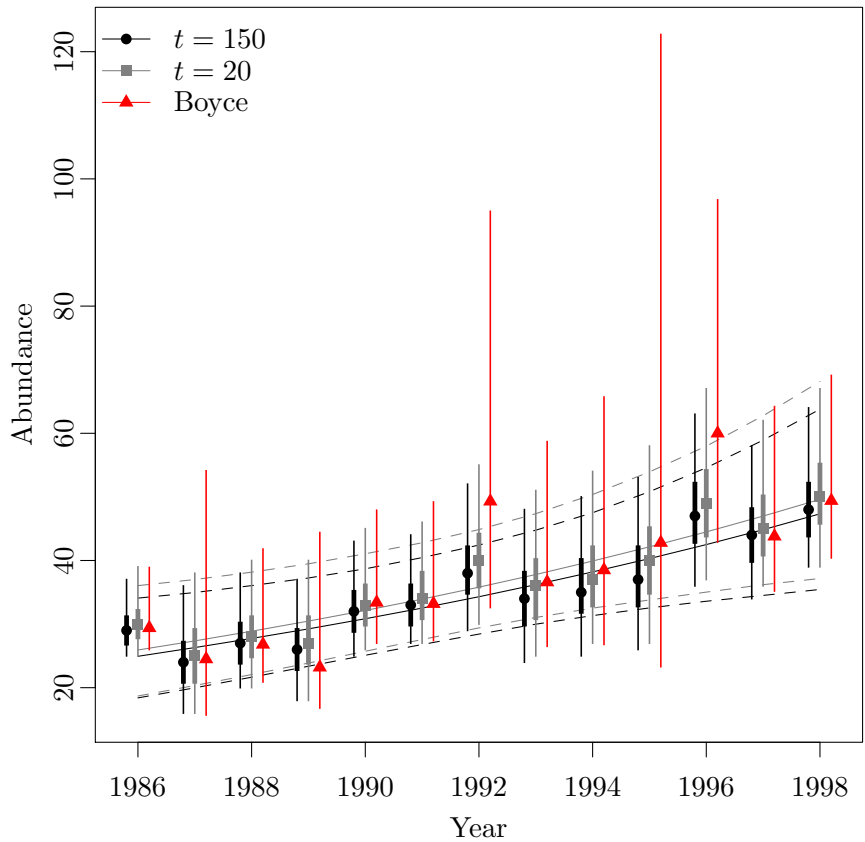

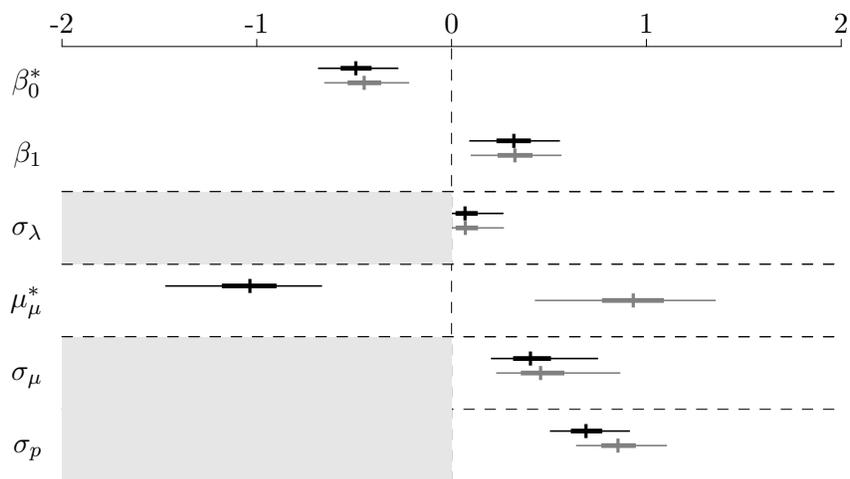
\begin{figure}[!htbp]
  \centering
  \input{grizzlypar}
  \caption{Posterior distribution estimates of parameters relating to abundance of female grizzly bears in the Yellowstone
ecosystem between 1986--1998 from fitting model \Mh to the data of Boyce et. al
(2001). The thin horizontal lines span the central 95\% credible interval, the thick horizontal line span the quartiles  of
the posterior distribution and the vertical line corresponds to the median.  The model fitted to $t=150$ is denoted in black, and $t=20$ denoted in gray.  A solid gray area is an inadmissible region in the parameter space.  The value $\beta_{0}^{*}
= \beta_{0} - 4$ and $\mu_{\mu}^{*} = \mu_{\mu} + 4$.}  \label{fig:grizzpar}
  
\end{figure}

For inference we used three chains of length 10,000 each with a tuning/burn-in phase of length 5,000. 
The value of $t$ (20 or 150) has little impact on the prediction of $N$ and the resulting trend (Figure \ref{fig:NGrizz}), but does, as expected, have a large impact on the parameter estimates related to the capture probability random effect (Figure  \ref{fig:grizzpar}).  A substantial amount of mass in the posterior distribution near $0$ for $\sigma_{\lambda}$ suggests that there is little evidence of overdispersion in the Poisson model.

The yearly estimates obtained largely agree with those obtained by
\cite{Boyce2001}.  The major difference between the estimates (and corresponding
uncertainties) of \cite{Boyce2001} and ours is due to the partial pooling
towards the Poisson regression model evident in our results.  Such pooling
modestly shrinks the uncertainty of most estimates, with large shrinkage evident
for estimates with high uncertainty in \cite{Boyce2001}.  Shrinkage in the
estimates is also evident in 1992 when the estimate from \cite{Boyce2001} is
high relative to the estimates two years either side.

Evidence for an increasing population of grizzly bear females over time is compelling: a 95\% credible interval for the mean increase in total female grizzlies from 1986 to 1998 is ($7.4,45.6$) for $t=20$ or ($6.7,41.0$) for $t=150$.  This corresponds to a 95\% credible interval for the relative population increase of ($22.5\%,206.4\%$) for $t=20$ and ($21.5\%,201.5\%$) for $t=150$ over the duration of the study (Figure \ref{fig:NGrizz}).

\subsubsection{Computational Comparison}
Above we have compared the different parameterizations and their implementation
theoretically.  Here we explore the practical differences in the MCMC algorithms
for CDL-$N$ and CDL-$w$.  In particular, we fit the model for the
grizzly bears mentioned above in JAGS both for CDL-$N$ (as described above) and
CDL-$w$ (as described in section \ref{sec:cdlw} and the supplementary materials
section 5).  To evaluate performance of the two approaches 
we look at (i) the Gelman diagnostic, $\hat{R}$ \cite[]{Brooks1998a}, (ii) the time taken to run the algorithm and (iii) the effective sample size for the parameters.  Both (i) and (iii) were computed using the {\tt coda} package in R.  These measures are not a definitive measure of performance, as how we express the models, as well as the specific algorithms and programming language used can have a dramatic effect on both the convergence and running time of the algorithms.

For this example, the CDL-$w$ model has larger effective sample sizes in general (Table \ref{tab:effss}).  However, once we incorporate the time required to run the algorithm (on the lead authors desktop machine the CDL-$N$ model took $\sim$30 minutes vs $\sim$50 minutes for the CDL-$w$ model), the CDL-$N$ model has better efficiency than the CDL-$w$ model.  The effective sample size per minute is higher for every parameter we considered when using CDL-$N$.


\section{Discussion} \label{sec:discussion}
Here we have shown how the CDL can be explicitly expressed in terms of the
abundance parameter $N$ for the capture recapture model M$_{\text{h}}$ in a
manner that is straight-forward to implement using generic Bayesian hierarchical
modeling software such as BUGS.  In particular, the trans-dimensional nature of
the model is no major impediment, and offers an advantage over the RDL approach
in that it facilitates hierarchical modeling of $N$. Although we have shown how
the RDL model can be modified to allow hierarchical modeling of $N$ this is not
straight-forward owing to the fact that $N$ is a derived quantity in the
CDL-$w$ model. 

Another way to fit these models that we have not discussed in detail
is to integrate any latent random variables (including those whose dimension is
defined by ${N_k}$) out of the likelihood.  In general, we are unable to
explicitly evaluate the necessary integrals.  Numerical integration in the form
of quadrature is an attractive alternative for models where the integral is of
low dimension.  In particular, finding the MLE using quadrature is often faster than obtaining an appropriately sized sample from
the posterior distribution using MCMC and has the advantage of well defined
convergence conditions \cite[]{Gimenez2010}.  The problem is that quadrature gets
extremely difficult (and resource hungry) as we begin to explore hierarchical
models that require high dimensional integrals.  This is the case for the
hierarchical models we consider here that extend model \Mh to describe data from
replicated studies with distinct cohorts. 

In discussing use of DA for the CDL-$w$ model, RDL emphasize the distinction between their data augmentation scheme and use of TD methods.  Similar commentary is subsequently provided by \cite{Royle2012}.  In particular, RDL argue that standard approaches for fitting CDL-$N$ require specialized TD algorithms and treat the problem as one of model-selection. They then state that the model selection aspect is not a feature of the DA approach used to fit the CDL-$w$ model and go on to characterize TD approaches as being difficult and inaccessible \cite[]{Royle2012}.

As we have described above (and shown in the supplementary materials), the DA model fitting scheme of RDL corresponds to a specific TD implementation.  This means that  
the problem of moving between models does exist in the CDL-$w$ formulation.  In presenting this argument we do not claim that their method is useful only because it is a special case of a TD algorithm. We are arguing that their characterization of TD algorithms and model selection for fitting \Mh as being
difficult,  inaccessible and unnecessary is without foundation.  In spite of a different underlying motivation, RDL have used an equivalent method in their solution to the problem.

The close relationship between the RDL approach and more usual TD formulations is hardly surprising in that \cite{Godsill2001} has shown that algorithms such as RJMCMC and the algorithm of \cite{Carlin1995} are all special cases of a general class of methods based on a product-space representation of the TD problem.  In this product-space representation, the TD problem is solved by exploring an expanded parameter space that is of fixed dimension and that includes parameters in the likelihood for each model as well as any supplemental variables.  The RDL model fits into this class of problems as we have illustrated in the supplementary materials.

A key difference between the RDL approach and more typical TD algorithms is the reparameterization from $N$ to the nonidentifiable parameter $\bm w$.  The practical consequence of this in terms of fitting \Mh is the introduction of a binomial prior on $N$.  Importantly, the observed data likelihood is unchanged; the reparameterization of the model does not change its description of the data other than through the introduction of this prior \cite[]{Gelman2004c}.

The product Bernoulli prior on $\bm w$ corresponds to an induced binomial prior on $N$. But $N$ is the parameter of fundamental interest in closed-population capture recapture problems.  While computationally convenient, reparameterization in terms of $\bm w$ has the unfortunate effect of relegating $N$ to the status of a derived parameter.  As we have showed here, while still possible to model in terms of $N$, the specification of such a model requires additional effort and care to ensure it is appropriate.  
The natural focus of closed-population mark-recapture modeling is on $N$ and it would seem desirable to allow hierarchical modeling explicitly in terms of $N$.

While we have focused on model \Mh, we are able to extend these ideas to more
complex closed population models.  The CDL-$N$ representation can be
used for any capture-recapture model that can be fit using CDL-$w$.  For
example, we could use a the CDL-$N$ representation when including individual
covariates \cite[for example][]{Royle2008a,Schofield2011}.

For one such extension, consider the data of \cite{Hook1980}, who estimated the number of people born in upstate New York between 1969 and 1974 with spina bifida by comparing recorded cases found on three ``lists'': birth certificates, death certificates and medical rehabilitation records.  
Two features of the study are relevant here.  Firstly, a quantity of interest is
the trend in spina bifida prevalence.  In particular, \cite{Hook1980} discuss
the apparent ``temporal decrease in [spina bifida prevalence] rates with birth
year'' after considering the yearly prevalence estimates.  Secondly, there
appear to be differences in dependencies among the lists \cite[]{Hook1980,
Madigan1997} that limit a simple Rasch-type implementation of model M$_\text{h}$
or M$_\text{th}$ \cite[]{Agresti1994,Bartolucci2001,Stanghellini2004,Bartolucci2006} for multiple list data.  In the
supplementary materials, section 6, we show how we can use a latent class model
\cite[]{Goodman1974,Bartolucci2003} for this multilist capture-recapture
dataset.  The model we consider includes a hierarchical model for spina bifida
prevalence that allows us to formally explore potential trends through time.   

The ability to model abundance in multiple closed population studies as illustrated here provides an alternative to modeling using traditional robust design models \cite[]{Pollock1982}.  While robust design models make full use of the data, there are many situations where complicated models are required in order to adequately describe components of abundance change such as survival and recruitment, e.g. \cite{Bailey2010}.  Provided we are (i) willing to accept a small loss in efficiency, and (ii) only interested in modeling the changes in abundance, the approach outlined here may provide a simpler alternative for modeling such data.



\section{acknowledgements}
MRS was partially funded by NSF Grants \#0934516 and \#0814194.

\bibliographystyle{asa}
\bibliography{ees}

\newpage

%

\begin{table}[htbp]
\begin{minipage}{\textwidth}
\begin{center}
  \begin{tabular}{crrrrr}
 & & \multicolumn{2}{c}{ESS} & \multicolumn{2}{c}{ESS/min}\\
 & $\hat{R}$ & Total & $\frac{\text{CDL-}N}{\text{CDL-}w}$ & Total & $\frac{\text{CDL-}N}{\text{CDL-}w}$ \\\hline
\multirow{2}{*}{$\beta_{0}$}	&	1.01	&	916	&	\multirow{2}{*}{1.00}	&	32.3	&	\multirow{2}{*}{1.73}	\\
	&	1.02	&	918	&		&	18.7	&		\\\hline
\multirow{2}{*}{$\beta_{1}$}	&	1.00	&	3295	&	\multirow{2}{*}{0.74}	&	116.3	&	\multirow{2}{*}{1.29}	\\
	&	1.00	&	4446	&		&	90.4	&		\\\hline
\multirow{2}{*}{$N_{1}$}	&	1.01	&	1604	&	\multirow{2}{*}{0.75}	&	56.6	&	\multirow{2}{*}{1.31}	\\
	&	1.02	&	2126	&		&	43.2	&		\\\hline
\multirow{2}{*}{$N_{2}$}	&	1.00	&	2449	&	\multirow{2}{*}{0.89}	&	86.4	&	\multirow{2}{*}{1.55}	\\
	&	1.00	&	2742	&		&	55.7	&		\\\hline
\multirow{2}{*}{$N_{3}$}	&	1.01	&	2041	&	\multirow{2}{*}{0.87}	&	72.0	&	\multirow{2}{*}{1.51}	\\
	&	1.01	&	2352	&		&	47.8	&		\\\hline
\multirow{2}{*}{$N_{4}$}	&	1.01	&	1794	&	\multirow{2}{*}{0.83}	&	63.3	&	\multirow{2}{*}{1.44}	\\
	&	1.01	&	2165	&		&	44.0	&		\\\hline
\multirow{2}{*}{$N_{5}$}	&	1.00	&	2065	&	\multirow{2}{*}{0.79}	&	72.9	&	\multirow{2}{*}{1.36}	\\
	&	1.01	&	2629	&		&	53.4	&		\\\hline
\multirow{2}{*}{$N_{6}$}	&	1.01	&	1428	&	\multirow{2}{*}{0.72}	&	50.4	&	\multirow{2}{*}{1.25}	\\
	&	1.01	&	1989	&		&	40.4	&		\\\hline
\multirow{2}{*}{$N_{7}$}	&	1.01	&	2285	&	\multirow{2}{*}{0.90}	&	80.6	&	\multirow{2}{*}{1.56}	\\
	&	1.00	&	2548	&		&	51.8	&		\\\hline
\multirow{2}{*}{$N_{8}$}	&	1.00	&	2269	&	\multirow{2}{*}{1.01}	&	80.1	&	\multirow{2}{*}{1.76}	\\
	&	1.01	&	2243	&		&	45.6	&		\\\hline
\multirow{2}{*}{$N_{9}$}	&	1.01	&	1900	&	\multirow{2}{*}{1.12}	&	67.0	&	\multirow{2}{*}{1.95}	\\
	&	1.01	&	1695	&		&	34.5	&		\\\hline
\multirow{2}{*}{$N_{10}$}	&	1.01	&	2173	&	\multirow{2}{*}{1.00}	&	76.7	&	\multirow{2}{*}{1.74}	\\
	&	1.00	&	2172	&		&	44.1	&		\\\hline
\multirow{2}{*}{$N_{11}$}	&	1.01	&	1618	&	\multirow{2}{*}{0.89}	&	57.1	&	\multirow{2}{*}{1.54}	\\
	&	1.00	&	1822	&		&	37.0	&		\\\hline
\multirow{2}{*}{$N_{12}$}	&	1.00	&	1552	&	\multirow{2}{*}{0.91}	&	54.8	&	\multirow{2}{*}{1.59}	\\
	&	1.01	&	1699	&		&	34.5	&		\\\hline
\multirow{2}{*}{$N_{13}$}	&	1.01	&	1255	&	\multirow{2}{*}{0.80}	&	44.3	&	\multirow{2}{*}{1.39}	\\
	&	1.01	&	1568	&		&	31.9	&		\\\hline
\multirow{2}{*}{$\sigma_{N}$}	&	1.00	&	3734	&	\multirow{2}{*}{0.86}	&	131.8	&	\multirow{2}{*}{1.49}	\\
	&	1.00	&	4360	&		&	88.6	&		\\\hline
\multirow{2}{*}{$\mu_{\mu}$}	&	1.02	&	526	&	\multirow{2}{*}{0.97}	&	18.6	&	\multirow{2}{*}{1.68}	\\
	&	1.04	&	545	&		&	11.1	&		\\\hline
\multirow{2}{*}{$\sigma_{\mu}$}	&	1.02	&	1515	&	\multirow{2}{*}{0.90}	&	53.5	&	\multirow{2}{*}{1.57}	\\
	&	1.00	&	1677	&		&	34.1	&		\\\hline
\multirow{2}{*}{$\sigma_{p}$}	&	1.02	&	668	&	\multirow{2}{*}{0.88}	&	23.6	&	\multirow{2}{*}{1.53}	\\
	&	1.03	&	760	&		&	15.4	&		\\
 \end{tabular}
  \caption{The Gelman diagnostic $\hat{R}$, effective sample size (ESS) and ESS per minute (EES/min) for selected parameters in the hierarchical model for the CDL-$N$ model and the CDL-$w$ model.  For each parameter the top line refers to the CDL-$N$ model and the bottom line the CDL-$w$ model.  Columns four and six take the ratio of ESS and ESS/min for the CDL-$N$ and CDL-$w$ models.}\label{tab:effss}
\end{center}
\end{minipage}
\end{table}


%



\end{document}

%% file: grizzlycomp.tex
\begin{tikzpicture}[x=1pt,y=1pt]
\definecolor[named]{drawColor}{rgb}{0.00,0.00,0.00}
\definecolor[named]{fillColor}{rgb}{1.00,1.00,1.00}
\fill[color=fillColor,fill opacity=0.00,] (0,0) rectangle (338.22,338.22);
\begin{scope}
\path[clip] (  0.00,  0.00) rectangle (338.22,338.22);
\definecolor[named]{drawColor}{rgb}{0.00,0.00,0.00}

\draw[color=drawColor,line cap=round,line join=round,fill opacity=0.00,] ( 52.12, 36.00) -- (310.10, 36.00);

\draw[color=drawColor,line cap=round,line join=round,fill opacity=0.00,] ( 52.12, 36.00) -- ( 52.12, 30.44);

\draw[color=drawColor,line cap=round,line join=round,fill opacity=0.00,] ( 95.12, 36.00) -- ( 95.12, 30.44);

\draw[color=drawColor,line cap=round,line join=round,fill opacity=0.00,] (138.12, 36.00) -- (138.12, 30.44);

\draw[color=drawColor,line cap=round,line join=round,fill opacity=0.00,] (181.11, 36.00) -- (181.11, 30.44);

\draw[color=drawColor,line cap=round,line join=round,fill opacity=0.00,] (224.11, 36.00) -- (224.11, 30.44);

\draw[color=drawColor,line cap=round,line join=round,fill opacity=0.00,] (267.10, 36.00) -- (267.10, 30.44);

\draw[color=drawColor,line cap=round,line join=round,fill opacity=0.00,] (310.10, 36.00) -- (310.10, 30.44);

\node[color=drawColor,anchor=base,inner sep=0pt, outer sep=0pt, scale=  1.00] at ( 52.12, 15.60) {1986};

\node[color=drawColor,anchor=base,inner sep=0pt, outer sep=0pt, scale=  1.00] at ( 95.12, 15.60) {1988};

\node[color=drawColor,anchor=base,inner sep=0pt, outer sep=0pt, scale=  1.00] at (138.12, 15.60) {1990};

\node[color=drawColor,anchor=base,inner sep=0pt, outer sep=0pt, scale=  1.00] at (181.11, 15.60) {1992};

\node[color=drawColor,anchor=base,inner sep=0pt, outer sep=0pt, scale=  1.00] at (224.11, 15.60) {1994};

\node[color=drawColor,anchor=base,inner sep=0pt, outer sep=0pt, scale=  1.00] at (267.10, 15.60) {1996};

\node[color=drawColor,anchor=base,inner sep=0pt, outer sep=0pt, scale=  1.00] at (310.10, 15.60) {1998};

\draw[color=drawColor,line cap=round,line join=round,fill opacity=0.00,] ( 36.00, 56.66) -- ( 36.00,297.42);

\draw[color=drawColor,line cap=round,line join=round,fill opacity=0.00,] ( 36.00, 56.66) -- ( 30.44, 56.66);

\draw[color=drawColor,line cap=round,line join=round,fill opacity=0.00,] ( 36.00,104.81) -- ( 30.44,104.81);

\draw[color=drawColor,line cap=round,line join=round,fill opacity=0.00,] ( 36.00,152.96) -- ( 30.44,152.96);

\draw[color=drawColor,line cap=round,line join=round,fill opacity=0.00,] ( 36.00,201.11) -- ( 30.44,201.11);

\draw[color=drawColor,line cap=round,line join=round,fill opacity=0.00,] ( 36.00,249.27) -- ( 30.44,249.27);

\draw[color=drawColor,line cap=round,line join=round,fill opacity=0.00,] ( 36.00,297.42) -- ( 30.44,297.42);

\node[rotate= 90.00,color=drawColor,anchor=base,inner sep=0pt, outer sep=0pt, scale=  1.00] at ( 27.60, 56.66) {20};

\node[rotate= 90.00,color=drawColor,anchor=base,inner sep=0pt, outer sep=0pt, scale=  1.00] at ( 27.60,104.81) {40};

\node[rotate= 90.00,color=drawColor,anchor=base,inner sep=0pt, outer sep=0pt, scale=  1.00] at ( 27.60,152.96) {60};

\node[rotate= 90.00,color=drawColor,anchor=base,inner sep=0pt, outer sep=0pt, scale=  1.00] at ( 27.60,201.11) {80};

\node[rotate= 90.00,color=drawColor,anchor=base,inner sep=0pt, outer sep=0pt, scale=  1.00] at ( 27.60,249.27) {100};

\node[rotate= 90.00,color=drawColor,anchor=base,inner sep=0pt, outer sep=0pt, scale=  1.00] at ( 27.60,297.42) {120};

\draw[color=drawColor,line cap=round,line join=round,fill opacity=0.00,] ( 36.00, 36.00) --
	(326.22, 36.00) --
	(326.22,314.22) --
	( 36.00,314.22) --
	( 36.00, 36.00);
\end{scope}
\begin{scope}
\path[clip] (  0.00,  0.00) rectangle (338.22,338.22);
\definecolor[named]{drawColor}{rgb}{0.00,0.00,0.00}

\node[color=drawColor,anchor=base,inner sep=0pt, outer sep=0pt, scale=  1.00] at (181.11,  0.00) {Year};

\node[rotate= 90.00,color=drawColor,anchor=base,inner sep=0pt, outer sep=0pt, scale=  1.00] at ( 12.00,175.11) {Abundance};
\end{scope}
\begin{scope}
\path[clip] ( 36.00, 36.00) rectangle (326.22,314.22);
\definecolor[named]{drawColor}{rgb}{0.00,0.00,0.00}

\draw[color=drawColor,line cap=round,line join=round,fill opacity=0.00,] ( 52.12, 68.54) --
	( 73.62, 71.84) --
	( 95.12, 75.30) --
	(116.62, 78.92) --
	(138.12, 82.76) --
	(159.61, 86.79) --
	(181.11, 91.09) --
	(202.61, 95.67) --
	(224.11,100.48) --
	(245.61,105.53) --
	(267.10,110.83) --
	(288.60,116.46) --
	(310.10,122.47);

\draw[color=drawColor,dash pattern=on 4pt off 4pt ,line cap=round,line join=round,fill opacity=0.00,] ( 52.12, 52.80) --
	( 73.62, 56.60) --
	( 95.12, 60.63) --
	(116.62, 64.78) --
	(138.12, 68.93) --
	(159.61, 73.00) --
	(181.11, 76.95) --
	(202.61, 80.70) --
	(224.11, 83.94) --
	(245.61, 86.86) --
	(267.10, 89.38) --
	(288.60, 91.59) --
	(310.10, 93.84);

\draw[color=drawColor,dash pattern=on 4pt off 4pt ,line cap=round,line join=round,fill opacity=0.00,] ( 52.12, 90.57) --
	( 73.62, 92.68) --
	( 95.12, 95.30) --
	(116.62, 98.19) --
	(138.12,101.66) --
	(159.61,105.90) --
	(181.11,110.65) --
	(202.61,116.28) --
	(224.11,122.89) --
	(245.61,130.86) --
	(267.10,139.91) --
	(288.60,150.36) --
	(310.10,162.08);
\definecolor[named]{drawColor}{rgb}{0.50,0.50,0.50}

\draw[color=drawColor,line cap=round,line join=round,fill opacity=0.00,] ( 52.12, 70.93) --
	( 73.62, 74.37) --
	( 95.12, 78.02) --
	(116.62, 81.87) --
	(138.12, 85.89) --
	(159.61, 90.18) --
	(181.11, 94.72) --
	(202.61, 99.50) --
	(224.11,104.61) --
	(245.61,109.98) --
	(267.10,115.63) --
	(288.60,121.62) --
	(310.10,127.94);

\draw[color=drawColor,dash pattern=on 4pt off 4pt ,line cap=round,line join=round,fill opacity=0.00,] ( 52.12, 53.51) --
	( 73.62, 57.49) --
	( 95.12, 61.63) --
	(116.62, 65.98) --
	(138.12, 70.44) --
	(159.61, 74.89) --
	(181.11, 79.29) --
	(202.61, 83.23) --
	(224.11, 86.78) --
	(245.61, 89.87) --
	(267.10, 92.84) --
	(288.60, 95.65) --
	(310.10, 98.00);

\draw[color=drawColor,dash pattern=on 4pt off 4pt ,line cap=round,line join=round,fill opacity=0.00,] ( 52.12, 95.29) --
	( 73.62, 97.60) --
	( 95.12,100.39) --
	(116.62,103.55) --
	(138.12,107.35) --
	(159.61,111.52) --
	(181.11,116.51) --
	(202.61,122.53) --
	(224.11,129.78) --
	(245.61,138.35) --
	(267.10,148.09) --
	(288.60,159.59) --
	(310.10,172.56);
\definecolor[named]{drawColor}{rgb}{0.00,0.00,0.00}

\draw[color=drawColor,line width= 0.6pt,line cap=rect,line join=round,fill opacity=0.00,] ( 47.82, 68.70) --
	( 47.82, 97.59);
\definecolor[named]{drawColor}{rgb}{0.50,0.50,0.50}

\draw[color=drawColor,line width= 0.6pt,line cap=rect,line join=round,fill opacity=0.00,] ( 52.12, 68.70) --
	( 52.12,102.40);
\definecolor[named]{drawColor}{rgb}{1.00,0.00,0.00}

\draw[color=drawColor,line width= 0.6pt,line cap=rect,line join=round,fill opacity=0.00,] ( 56.42, 71.10) --
	( 56.42,102.16);
\definecolor[named]{drawColor}{rgb}{0.00,0.00,0.00}

\draw[color=drawColor,line width= 1.8pt,line cap=rect,line join=round,fill opacity=0.00,] ( 47.82, 73.51) --
	( 47.82, 83.14);
\definecolor[named]{drawColor}{rgb}{0.50,0.50,0.50}

\draw[color=drawColor,line width= 1.8pt,line cap=rect,line join=round,fill opacity=0.00,] ( 52.12, 75.92) --
	( 52.12, 85.55);
\definecolor[named]{fillColor}{rgb}{0.00,0.00,0.00}

\draw[fill=fillColor,draw opacity=0.00,] ( 47.82, 78.33) circle (  2.03);
\definecolor[named]{fillColor}{rgb}{0.50,0.50,0.50}

\draw[fill=fillColor,draw opacity=0.00,] ( 50.10, 78.71) --
	( 54.15, 78.71) --
	( 54.15, 82.76) --
	( 50.10, 82.76) --
	cycle;
\definecolor[named]{fillColor}{rgb}{1.00,0.00,0.00}

\draw[fill=fillColor,draw opacity=0.00,] ( 56.42, 82.44) --
	( 59.15, 77.71) --
	( 53.70, 77.71) --
	cycle;
\definecolor[named]{drawColor}{rgb}{0.00,0.00,0.00}

\draw[color=drawColor,line width= 0.6pt,line cap=rect,line join=round,fill opacity=0.00,] ( 69.32, 47.03) --
	( 69.32, 95.18);
\definecolor[named]{drawColor}{rgb}{0.50,0.50,0.50}

\draw[color=drawColor,line width= 0.6pt,line cap=rect,line join=round,fill opacity=0.00,] ( 73.62, 47.03) --
	( 73.62, 99.99);
\definecolor[named]{drawColor}{rgb}{1.00,0.00,0.00}

\draw[color=drawColor,line width= 0.6pt,line cap=rect,line join=round,fill opacity=0.00,] ( 77.92, 46.30) --
	( 77.92,138.76);
\definecolor[named]{drawColor}{rgb}{0.00,0.00,0.00}

\draw[color=drawColor,line width= 1.8pt,line cap=rect,line join=round,fill opacity=0.00,] ( 69.32, 59.06) --
	( 69.32, 73.51);
\definecolor[named]{drawColor}{rgb}{0.50,0.50,0.50}

\draw[color=drawColor,line width= 1.8pt,line cap=rect,line join=round,fill opacity=0.00,] ( 73.62, 59.06) --
	( 73.62, 78.33);
\definecolor[named]{fillColor}{rgb}{0.00,0.00,0.00}

\draw[fill=fillColor,draw opacity=0.00,] ( 69.32, 66.29) circle (  2.03);
\definecolor[named]{fillColor}{rgb}{0.50,0.50,0.50}

\draw[fill=fillColor,draw opacity=0.00,] ( 71.60, 66.67) --
	( 75.65, 66.67) --
	( 75.65, 70.72) --
	( 71.60, 70.72) --
	cycle;
\definecolor[named]{fillColor}{rgb}{1.00,0.00,0.00}

\draw[fill=fillColor,draw opacity=0.00,] ( 77.92, 70.64) --
	( 80.65, 65.92) --
	( 75.19, 65.92) --
	cycle;
\definecolor[named]{drawColor}{rgb}{0.00,0.00,0.00}

\draw[color=drawColor,line width= 0.6pt,line cap=rect,line join=round,fill opacity=0.00,] ( 90.82, 56.66) --
	( 90.82, 99.99);
\definecolor[named]{drawColor}{rgb}{0.50,0.50,0.50}

\draw[color=drawColor,line width= 0.6pt,line cap=rect,line join=round,fill opacity=0.00,] ( 95.12, 56.66) --
	( 95.12,104.81);
\definecolor[named]{drawColor}{rgb}{1.00,0.00,0.00}

\draw[color=drawColor,line width= 0.6pt,line cap=rect,line join=round,fill opacity=0.00,] ( 99.42, 58.82) --
	( 99.42,109.14);
\definecolor[named]{drawColor}{rgb}{0.00,0.00,0.00}

\draw[color=drawColor,line width= 1.8pt,line cap=rect,line join=round,fill opacity=0.00,] ( 90.82, 66.29) --
	( 90.82, 80.73);
\definecolor[named]{drawColor}{rgb}{0.50,0.50,0.50}

\draw[color=drawColor,line width= 1.8pt,line cap=rect,line join=round,fill opacity=0.00,] ( 95.12, 68.70) --
	( 95.12, 83.14);
\definecolor[named]{fillColor}{rgb}{0.00,0.00,0.00}

\draw[fill=fillColor,draw opacity=0.00,] ( 90.82, 73.51) circle (  2.03);
\definecolor[named]{fillColor}{rgb}{0.50,0.50,0.50}

\draw[fill=fillColor,draw opacity=0.00,] ( 93.09, 73.89) --
	( 97.14, 73.89) --
	( 97.14, 77.94) --
	( 93.09, 77.94) --
	cycle;
\definecolor[named]{fillColor}{rgb}{1.00,0.00,0.00}

\draw[fill=fillColor,draw opacity=0.00,] ( 99.42, 76.18) --
	(102.15, 71.45) --
	( 96.69, 71.45) --
	cycle;
\definecolor[named]{drawColor}{rgb}{0.00,0.00,0.00}

\draw[color=drawColor,line width= 0.6pt,line cap=rect,line join=round,fill opacity=0.00,] (112.32, 51.84) --
	(112.32, 97.59);
\definecolor[named]{drawColor}{rgb}{0.50,0.50,0.50}

\draw[color=drawColor,line width= 0.6pt,line cap=rect,line join=round,fill opacity=0.00,] (116.62, 51.84) --
	(116.62,104.81);
\definecolor[named]{drawColor}{rgb}{1.00,0.00,0.00}

\draw[color=drawColor,line width= 0.6pt,line cap=rect,line join=round,fill opacity=0.00,] (120.92, 48.95) --
	(120.92,115.40);
\definecolor[named]{drawColor}{rgb}{0.00,0.00,0.00}

\draw[color=drawColor,line width= 1.8pt,line cap=rect,line join=round,fill opacity=0.00,] (112.32, 63.88) --
	(112.32, 78.33);
\definecolor[named]{drawColor}{rgb}{0.50,0.50,0.50}

\draw[color=drawColor,line width= 1.8pt,line cap=rect,line join=round,fill opacity=0.00,] (116.62, 66.29) --
	(116.62, 83.14);
\definecolor[named]{fillColor}{rgb}{0.00,0.00,0.00}

\draw[fill=fillColor,draw opacity=0.00,] (112.32, 71.10) circle (  2.03);
\definecolor[named]{fillColor}{rgb}{0.50,0.50,0.50}

\draw[fill=fillColor,draw opacity=0.00,] (114.59, 71.49) --
	(118.64, 71.49) --
	(118.64, 75.54) --
	(114.59, 75.54) --
	cycle;
\definecolor[named]{fillColor}{rgb}{1.00,0.00,0.00}

\draw[fill=fillColor,draw opacity=0.00,] (120.92, 67.51) --
	(123.64, 62.79) --
	(118.19, 62.79) --
	cycle;
\definecolor[named]{drawColor}{rgb}{0.00,0.00,0.00}

\draw[color=drawColor,line width= 0.6pt,line cap=rect,line join=round,fill opacity=0.00,] (133.82, 68.70) --
	(133.82,112.03);
\definecolor[named]{drawColor}{rgb}{0.50,0.50,0.50}

\draw[color=drawColor,line width= 0.6pt,line cap=rect,line join=round,fill opacity=0.00,] (138.12, 71.10) --
	(138.12,116.85);
\definecolor[named]{drawColor}{rgb}{1.00,0.00,0.00}

\draw[color=drawColor,line width= 0.6pt,line cap=rect,line join=round,fill opacity=0.00,] (142.42, 73.51) --
	(142.42,123.83);
\definecolor[named]{drawColor}{rgb}{0.00,0.00,0.00}

\draw[color=drawColor,line width= 1.8pt,line cap=rect,line join=round,fill opacity=0.00,] (133.82, 78.33) --
	(133.82, 92.77);
\definecolor[named]{drawColor}{rgb}{0.50,0.50,0.50}

\draw[color=drawColor,line width= 1.8pt,line cap=rect,line join=round,fill opacity=0.00,] (138.12, 80.73) --
	(138.12, 95.18);
\definecolor[named]{fillColor}{rgb}{0.00,0.00,0.00}

\draw[fill=fillColor,draw opacity=0.00,] (133.82, 85.55) circle (  2.03);
\definecolor[named]{fillColor}{rgb}{0.50,0.50,0.50}

\draw[fill=fillColor,draw opacity=0.00,] (136.09, 85.93) --
	(140.14, 85.93) --
	(140.14, 89.98) --
	(136.09, 89.98) --
	cycle;
\definecolor[named]{fillColor}{rgb}{1.00,0.00,0.00}

\draw[fill=fillColor,draw opacity=0.00,] (142.42, 92.07) --
	(145.14, 87.34) --
	(139.69, 87.34) --
	cycle;
\definecolor[named]{drawColor}{rgb}{0.00,0.00,0.00}

\draw[color=drawColor,line width= 0.6pt,line cap=rect,line join=round,fill opacity=0.00,] (155.31, 73.51) --
	(155.31,114.44);
\definecolor[named]{drawColor}{rgb}{0.50,0.50,0.50}

\draw[color=drawColor,line width= 0.6pt,line cap=rect,line join=round,fill opacity=0.00,] (159.61, 73.51) --
	(159.61,119.26);
\definecolor[named]{drawColor}{rgb}{1.00,0.00,0.00}

\draw[color=drawColor,line width= 0.6pt,line cap=rect,line join=round,fill opacity=0.00,] (163.91, 74.47) --
	(163.91,126.96);
\definecolor[named]{drawColor}{rgb}{0.00,0.00,0.00}

\draw[color=drawColor,line width= 1.8pt,line cap=rect,line join=round,fill opacity=0.00,] (155.31, 80.73) --
	(155.31, 95.18);
\definecolor[named]{drawColor}{rgb}{0.50,0.50,0.50}

\draw[color=drawColor,line width= 1.8pt,line cap=rect,line join=round,fill opacity=0.00,] (159.61, 83.14) --
	(159.61, 99.99);
\definecolor[named]{fillColor}{rgb}{0.00,0.00,0.00}

\draw[fill=fillColor,draw opacity=0.00,] (155.31, 87.96) circle (  2.03);
\definecolor[named]{fillColor}{rgb}{0.50,0.50,0.50}

\draw[fill=fillColor,draw opacity=0.00,] (157.59, 88.34) --
	(161.64, 88.34) --
	(161.64, 92.39) --
	(157.59, 92.39) --
	cycle;
\definecolor[named]{fillColor}{rgb}{1.00,0.00,0.00}

\draw[fill=fillColor,draw opacity=0.00,] (163.91, 91.59) --
	(166.64, 86.86) --
	(161.19, 86.86) --
	cycle;
\definecolor[named]{drawColor}{rgb}{0.00,0.00,0.00}

\draw[color=drawColor,line width= 0.6pt,line cap=rect,line join=round,fill opacity=0.00,] (176.81, 78.33) --
	(176.81,133.70);
\definecolor[named]{drawColor}{rgb}{0.50,0.50,0.50}

\draw[color=drawColor,line width= 0.6pt,line cap=rect,line join=round,fill opacity=0.00,] (181.11, 80.73) --
	(181.11,140.92);
\definecolor[named]{drawColor}{rgb}{1.00,0.00,0.00}

\draw[color=drawColor,line width= 0.6pt,line cap=rect,line join=round,fill opacity=0.00,] (185.41, 86.99) --
	(185.41,236.99);
\definecolor[named]{drawColor}{rgb}{0.00,0.00,0.00}

\draw[color=drawColor,line width= 1.8pt,line cap=rect,line join=round,fill opacity=0.00,] (176.81, 92.77) --
	(176.81,109.62);
\definecolor[named]{drawColor}{rgb}{0.50,0.50,0.50}

\draw[color=drawColor,line width= 1.8pt,line cap=rect,line join=round,fill opacity=0.00,] (181.11, 95.18) --
	(181.11,114.44);
\definecolor[named]{fillColor}{rgb}{0.00,0.00,0.00}

\draw[fill=fillColor,draw opacity=0.00,] (176.81, 99.99) circle (  2.03);
\definecolor[named]{fillColor}{rgb}{0.50,0.50,0.50}

\draw[fill=fillColor,draw opacity=0.00,] (179.09,102.78) --
	(183.14,102.78) --
	(183.14,106.83) --
	(179.09,106.83) --
	cycle;
\definecolor[named]{fillColor}{rgb}{1.00,0.00,0.00}

\draw[fill=fillColor,draw opacity=0.00,] (185.41,130.35) --
	(188.14,125.63) --
	(182.68,125.63) --
	cycle;
\definecolor[named]{drawColor}{rgb}{0.00,0.00,0.00}

\draw[color=drawColor,line width= 0.6pt,line cap=rect,line join=round,fill opacity=0.00,] (198.31, 66.29) --
	(198.31,124.07);
\definecolor[named]{drawColor}{rgb}{0.50,0.50,0.50}

\draw[color=drawColor,line width= 0.6pt,line cap=rect,line join=round,fill opacity=0.00,] (202.61, 68.70) --
	(202.61,131.29);
\definecolor[named]{drawColor}{rgb}{1.00,0.00,0.00}

\draw[color=drawColor,line width= 0.6pt,line cap=rect,line join=round,fill opacity=0.00,] (206.91, 72.31) --
	(206.91,149.83);
\definecolor[named]{drawColor}{rgb}{0.00,0.00,0.00}

\draw[color=drawColor,line width= 1.8pt,line cap=rect,line join=round,fill opacity=0.00,] (198.31, 80.73) --
	(198.31, 99.99);
\definecolor[named]{drawColor}{rgb}{0.50,0.50,0.50}

\draw[color=drawColor,line width= 1.8pt,line cap=rect,line join=round,fill opacity=0.00,] (202.61, 83.14) --
	(202.61,104.81);
\definecolor[named]{fillColor}{rgb}{0.00,0.00,0.00}

\draw[fill=fillColor,draw opacity=0.00,] (198.31, 90.36) circle (  2.03);
\definecolor[named]{fillColor}{rgb}{0.50,0.50,0.50}

\draw[fill=fillColor,draw opacity=0.00,] (200.58, 93.15) --
	(204.63, 93.15) --
	(204.63, 97.20) --
	(200.58, 97.20) --
	cycle;
\definecolor[named]{fillColor}{rgb}{1.00,0.00,0.00}

\draw[fill=fillColor,draw opacity=0.00,] (206.91, 99.77) --
	(209.64, 95.05) --
	(204.18, 95.05) --
	cycle;
\definecolor[named]{drawColor}{rgb}{0.00,0.00,0.00}

\draw[color=drawColor,line width= 0.6pt,line cap=rect,line join=round,fill opacity=0.00,] (219.81, 68.70) --
	(219.81,128.89);
\definecolor[named]{drawColor}{rgb}{0.50,0.50,0.50}

\draw[color=drawColor,line width= 0.6pt,line cap=rect,line join=round,fill opacity=0.00,] (224.11, 73.51) --
	(224.11,138.52);
\definecolor[named]{drawColor}{rgb}{1.00,0.00,0.00}

\draw[color=drawColor,line width= 0.6pt,line cap=rect,line join=round,fill opacity=0.00,] (228.41, 73.03) --
	(228.41,166.69);
\definecolor[named]{drawColor}{rgb}{0.00,0.00,0.00}

\draw[color=drawColor,line width= 1.8pt,line cap=rect,line join=round,fill opacity=0.00,] (219.81, 85.55) --
	(219.81,104.81);
\definecolor[named]{drawColor}{rgb}{0.50,0.50,0.50}

\draw[color=drawColor,line width= 1.8pt,line cap=rect,line join=round,fill opacity=0.00,] (224.11, 87.96) --
	(224.11,109.62);
\definecolor[named]{fillColor}{rgb}{0.00,0.00,0.00}

\draw[fill=fillColor,draw opacity=0.00,] (219.81, 92.77) circle (  2.03);
\definecolor[named]{fillColor}{rgb}{0.50,0.50,0.50}

\draw[fill=fillColor,draw opacity=0.00,] (222.08, 95.56) --
	(226.13, 95.56) --
	(226.13, 99.61) --
	(222.08, 99.61) --
	cycle;
\definecolor[named]{fillColor}{rgb}{1.00,0.00,0.00}

\draw[fill=fillColor,draw opacity=0.00,] (228.41,104.35) --
	(231.13, 99.62) --
	(225.68, 99.62) --
	cycle;
\definecolor[named]{drawColor}{rgb}{0.00,0.00,0.00}

\draw[color=drawColor,line width= 0.6pt,line cap=rect,line join=round,fill opacity=0.00,] (241.31, 71.10) --
	(241.31,136.11);
\definecolor[named]{drawColor}{rgb}{0.50,0.50,0.50}

\draw[color=drawColor,line width= 0.6pt,line cap=rect,line join=round,fill opacity=0.00,] (245.61, 73.51) --
	(245.61,148.15);
\definecolor[named]{drawColor}{rgb}{1.00,0.00,0.00}

\draw[color=drawColor,line width= 0.6pt,line cap=rect,line join=round,fill opacity=0.00,] (249.91, 64.60) --
	(249.91,303.92);
\definecolor[named]{drawColor}{rgb}{0.00,0.00,0.00}

\draw[color=drawColor,line width= 1.8pt,line cap=rect,line join=round,fill opacity=0.00,] (241.31, 87.96) --
	(241.31,109.62);
\definecolor[named]{drawColor}{rgb}{0.50,0.50,0.50}

\draw[color=drawColor,line width= 1.8pt,line cap=rect,line join=round,fill opacity=0.00,] (245.61, 92.77) --
	(245.61,116.85);
\definecolor[named]{fillColor}{rgb}{0.00,0.00,0.00}

\draw[fill=fillColor,draw opacity=0.00,] (241.31, 97.59) circle (  2.03);
\definecolor[named]{fillColor}{rgb}{0.50,0.50,0.50}

\draw[fill=fillColor,draw opacity=0.00,] (243.58,102.78) --
	(247.63,102.78) --
	(247.63,106.83) --
	(243.58,106.83) --
	cycle;
\definecolor[named]{fillColor}{rgb}{1.00,0.00,0.00}

\draw[fill=fillColor,draw opacity=0.00,] (249.91,114.70) --
	(252.63,109.98) --
	(247.18,109.98) --
	cycle;
\definecolor[named]{drawColor}{rgb}{0.00,0.00,0.00}

\draw[color=drawColor,line width= 0.6pt,line cap=rect,line join=round,fill opacity=0.00,] (262.80, 95.18) --
	(262.80,160.18);
\definecolor[named]{drawColor}{rgb}{0.50,0.50,0.50}

\draw[color=drawColor,line width= 0.6pt,line cap=rect,line join=round,fill opacity=0.00,] (267.10, 97.59) --
	(267.10,169.82);
\definecolor[named]{drawColor}{rgb}{1.00,0.00,0.00}

\draw[color=drawColor,line width= 0.6pt,line cap=rect,line join=round,fill opacity=0.00,] (271.40,111.79) --
	(271.40,241.32);
\definecolor[named]{drawColor}{rgb}{0.00,0.00,0.00}

\draw[color=drawColor,line width= 1.8pt,line cap=rect,line join=round,fill opacity=0.00,] (262.80,112.03) --
	(262.80,133.70);
\definecolor[named]{drawColor}{rgb}{0.50,0.50,0.50}

\draw[color=drawColor,line width= 1.8pt,line cap=rect,line join=round,fill opacity=0.00,] (267.10,114.44) --
	(267.10,138.52);
\definecolor[named]{fillColor}{rgb}{0.00,0.00,0.00}

\draw[fill=fillColor,draw opacity=0.00,] (262.80,121.66) circle (  2.03);
\definecolor[named]{fillColor}{rgb}{0.50,0.50,0.50}

\draw[fill=fillColor,draw opacity=0.00,] (265.08,124.45) --
	(269.13,124.45) --
	(269.13,128.50) --
	(265.08,128.50) --
	cycle;
\definecolor[named]{fillColor}{rgb}{1.00,0.00,0.00}

\draw[fill=fillColor,draw opacity=0.00,] (271.40,156.11) --
	(274.13,151.39) --
	(268.68,151.39) --
	cycle;
\definecolor[named]{drawColor}{rgb}{0.00,0.00,0.00}

\draw[color=drawColor,line width= 0.6pt,line cap=rect,line join=round,fill opacity=0.00,] (284.30, 90.36) --
	(284.30,148.15);
\definecolor[named]{drawColor}{rgb}{0.50,0.50,0.50}

\draw[color=drawColor,line width= 0.6pt,line cap=rect,line join=round,fill opacity=0.00,] (288.60, 95.18) --
	(288.60,157.78);
\definecolor[named]{drawColor}{rgb}{1.00,0.00,0.00}

\draw[color=drawColor,line width= 0.6pt,line cap=rect,line join=round,fill opacity=0.00,] (292.90, 93.25) --
	(292.90,163.07);
\definecolor[named]{drawColor}{rgb}{0.00,0.00,0.00}

\draw[color=drawColor,line width= 1.8pt,line cap=rect,line join=round,fill opacity=0.00,] (284.30,104.81) --
	(284.30,124.07);
\definecolor[named]{drawColor}{rgb}{0.50,0.50,0.50}

\draw[color=drawColor,line width= 1.8pt,line cap=rect,line join=round,fill opacity=0.00,] (288.60,107.22) --
	(288.60,128.89);
\definecolor[named]{fillColor}{rgb}{0.00,0.00,0.00}

\draw[fill=fillColor,draw opacity=0.00,] (284.30,114.44) circle (  2.03);
\definecolor[named]{fillColor}{rgb}{0.50,0.50,0.50}

\draw[fill=fillColor,draw opacity=0.00,] (286.58,114.82) --
	(290.63,114.82) --
	(290.63,118.87) --
	(286.58,118.87) --
	cycle;
\definecolor[named]{fillColor}{rgb}{1.00,0.00,0.00}

\draw[fill=fillColor,draw opacity=0.00,] (292.90,117.11) --
	(295.63,112.38) --
	(290.17,112.38) --
	cycle;
\definecolor[named]{drawColor}{rgb}{0.00,0.00,0.00}

\draw[color=drawColor,line width= 0.6pt,line cap=rect,line join=round,fill opacity=0.00,] (305.80,102.40) --
	(305.80,162.59);
\definecolor[named]{drawColor}{rgb}{0.50,0.50,0.50}

\draw[color=drawColor,line width= 0.6pt,line cap=rect,line join=round,fill opacity=0.00,] (310.10,102.40) --
	(310.10,169.82);
\definecolor[named]{drawColor}{rgb}{1.00,0.00,0.00}

\draw[color=drawColor,line width= 0.6pt,line cap=rect,line join=round,fill opacity=0.00,] (314.40,105.77) --
	(314.40,174.87);
\definecolor[named]{drawColor}{rgb}{0.00,0.00,0.00}

\draw[color=drawColor,line width= 1.8pt,line cap=rect,line join=round,fill opacity=0.00,] (305.80,114.44) --
	(305.80,133.70);
\definecolor[named]{drawColor}{rgb}{0.50,0.50,0.50}

\draw[color=drawColor,line width= 1.8pt,line cap=rect,line join=round,fill opacity=0.00,] (310.10,119.26) --
	(310.10,140.92);
\definecolor[named]{fillColor}{rgb}{0.00,0.00,0.00}

\draw[fill=fillColor,draw opacity=0.00,] (305.80,124.07) circle (  2.03);
\definecolor[named]{fillColor}{rgb}{0.50,0.50,0.50}

\draw[fill=fillColor,draw opacity=0.00,] (308.08,126.86) --
	(312.13,126.86) --
	(312.13,130.91) --
	(308.08,130.91) --
	cycle;
\definecolor[named]{fillColor}{rgb}{1.00,0.00,0.00}

\draw[fill=fillColor,draw opacity=0.00,] (314.40,130.59) --
	(317.13,125.87) --
	(311.67,125.87) --
	cycle;
\definecolor[named]{drawColor}{rgb}{0.00,0.00,0.00}

\draw[color=drawColor,line cap=round,line join=round,fill opacity=0.00,] ( 38.70,302.22) -- ( 56.70,302.22);
\definecolor[named]{drawColor}{rgb}{0.50,0.50,0.50}

\draw[color=drawColor,line cap=round,line join=round,fill opacity=0.00,] ( 38.70,290.22) -- ( 56.70,290.22);
\definecolor[named]{drawColor}{rgb}{1.00,0.00,0.00}

\draw[color=drawColor,line cap=round,line join=round,fill opacity=0.00,] ( 38.70,278.22) -- ( 56.70,278.22);
\definecolor[named]{fillColor}{rgb}{0.00,0.00,0.00}

\draw[fill=fillColor,draw opacity=0.00,] ( 47.70,302.22) circle (  2.03);
\definecolor[named]{fillColor}{rgb}{0.50,0.50,0.50}

\draw[fill=fillColor,draw opacity=0.00,] ( 45.68,288.20) --
	( 49.73,288.20) --
	( 49.73,292.25) --
	( 45.68,292.25) --
	cycle;
\definecolor[named]{fillColor}{rgb}{1.00,0.00,0.00}

\draw[fill=fillColor,draw opacity=0.00,] ( 47.70,281.37) --
	( 50.43,276.65) --
	( 44.97,276.65) --
	cycle;
\definecolor[named]{drawColor}{rgb}{0.00,0.00,0.00}

\node[color=drawColor,anchor=base west,inner sep=0pt, outer sep=0pt, scale=  1.00] at ( 65.70,298.78) {$t=150$};

\node[color=drawColor,anchor=base west,inner sep=0pt, outer sep=0pt, scale=  1.00] at ( 65.70,286.78) {$t=20$};

\node[color=drawColor,anchor=base west,inner sep=0pt, outer sep=0pt, scale=  1.00] at ( 65.70,274.78) {Boyce};
\end{scope}
\end{tikzpicture}

%% file: grizzlypar.tex
\begin{tikzpicture}[x=1pt,y=1pt]
\definecolor[named]{drawColor}{rgb}{0.00,0.00,0.00}
\definecolor[named]{fillColor}{rgb}{1.00,1.00,1.00}
\fill[color=fillColor,fill opacity=0.00,] (0,0) rectangle (338.22,216.81);
\begin{scope}
\path[clip] (  0.00,  0.00) rectangle (338.22,192.81);
\end{scope}
\begin{scope}
\path[clip] (  0.00,  0.00) rectangle (338.22,192.81);
\definecolor[named]{drawColor}{rgb}{0.00,0.00,0.00}

\node[color=drawColor,anchor=base east,inner sep=0pt, outer sep=0pt, scale=  1.00] at ( 23.58,155.70) {$\beta_{0}^{*}$};

\draw[color=drawColor,line width= 0.6pt,line cap=rect,line join=round,fill opacity=0.00,] (128.12,163.70) --
	(157.93,163.70);

\draw[color=drawColor,line width= 1.8pt,line cap=rect,line join=round,fill opacity=0.00,] (137.24,163.70) --
	(147.17,163.70);

\draw[color=drawColor,line width= 1.2pt,line cap=rect,line join=round,fill opacity=0.00,] (142.10,165.76) --
	(142.10,161.64);
\definecolor[named]{drawColor}{rgb}{0.50,0.50,0.50}

\draw[color=drawColor,line width= 0.6pt,line cap=rect,line join=round,fill opacity=0.00,] (130.46,158.20) --
	(162.00,158.20);

\draw[color=drawColor,line width= 1.8pt,line cap=rect,line join=round,fill opacity=0.00,] (139.91,158.20) --
	(150.80,158.20);

\draw[color=drawColor,line width= 1.2pt,line cap=rect,line join=round,fill opacity=0.00,] (145.27,160.26) --
	(145.27,156.14);
\definecolor[named]{drawColor}{rgb}{0.00,0.00,0.00}

\node[color=drawColor,anchor=base east,inner sep=0pt, outer sep=0pt, scale=  1.00] at ( 23.58,128.24) {$\beta_{1}$};

\draw[color=drawColor,line width= 0.6pt,line cap=rect,line join=round,fill opacity=0.00,] (185.31,136.23) --
	(219.01,136.23);

\draw[color=drawColor,line width= 1.8pt,line cap=rect,line join=round,fill opacity=0.00,] (196.20,136.23) --
	(207.41,136.23);

\draw[color=drawColor,line width= 1.2pt,line cap=rect,line join=round,fill opacity=0.00,] (201.87,138.29) --
	(201.87,134.17);
\definecolor[named]{drawColor}{rgb}{0.50,0.50,0.50}

\draw[color=drawColor,line width= 0.6pt,line cap=rect,line join=round,fill opacity=0.00,] (185.83,130.74) --
	(219.64,130.74);

\draw[color=drawColor,line width= 1.8pt,line cap=rect,line join=round,fill opacity=0.00,] (196.61,130.74) --
	(208.05,130.74);

\draw[color=drawColor,line width= 1.2pt,line cap=rect,line join=round,fill opacity=0.00,] (202.27,132.80) --
	(202.27,128.68);
\definecolor[named]{drawColor}{rgb}{0.00,0.00,0.00}

\node[color=drawColor,anchor=base east,inner sep=0pt, outer sep=0pt, scale=  1.00] at ( 23.58,100.77) {$\sigma_{\lambda}$};

\draw[color=drawColor,line width= 0.6pt,line cap=rect,line join=round,fill opacity=0.00,] (178.56,108.76) --
	(197.65,108.76);

\draw[color=drawColor,line width= 1.8pt,line cap=rect,line join=round,fill opacity=0.00,] (180.72,108.76) --
	(187.29,108.76);

\draw[color=drawColor,line width= 1.2pt,line cap=rect,line join=round,fill opacity=0.00,] (183.41,110.82) --
	(183.41,106.70);
\definecolor[named]{drawColor}{rgb}{0.50,0.50,0.50}

\draw[color=drawColor,line width= 0.6pt,line cap=rect,line join=round,fill opacity=0.00,] (178.55,103.27) --
	(197.83,103.27);

\draw[color=drawColor,line width= 1.8pt,line cap=rect,line join=round,fill opacity=0.00,] (180.76,103.27) --
	(187.44,103.27);

\draw[color=drawColor,line width= 1.2pt,line cap=rect,line join=round,fill opacity=0.00,] (183.53,105.33) --
	(183.53,101.21);
\definecolor[named]{drawColor}{rgb}{0.00,0.00,0.00}

\node[color=drawColor,anchor=base east,inner sep=0pt, outer sep=0pt, scale=  1.00] at ( 23.58, 73.31) {$\mu_{\mu}^{*}$};

\draw[color=drawColor,line width= 0.6pt,line cap=rect,line join=round,fill opacity=0.00,] ( 70.35, 81.30) --
	(129.02, 81.30);

\draw[color=drawColor,line width= 1.8pt,line cap=rect,line join=round,fill opacity=0.00,] ( 92.33, 81.30) --
	(111.23, 81.30);

\draw[color=drawColor,line width= 1.2pt,line cap=rect,line join=round,fill opacity=0.00,] (102.05, 83.36) --
	(102.05, 79.24);
\definecolor[named]{drawColor}{rgb}{0.50,0.50,0.50}

\draw[color=drawColor,line width= 0.6pt,line cap=rect,line join=round,fill opacity=0.00,] (210.12, 75.81) --
	(277.90, 75.81);

\draw[color=drawColor,line width= 1.8pt,line cap=rect,line join=round,fill opacity=0.00,] (236.12, 75.81) --
	(257.76, 75.81);

\draw[color=drawColor,line width= 1.2pt,line cap=rect,line join=round,fill opacity=0.00,] (247.05, 77.87) --
	(247.05, 73.75);
\definecolor[named]{drawColor}{rgb}{0.00,0.00,0.00}

\node[color=drawColor,anchor=base east,inner sep=0pt, outer sep=0pt, scale=  1.00] at ( 23.58, 45.84) {$\sigma_{\mu}$};

\draw[color=drawColor,line width= 0.6pt,line cap=rect,line join=round,fill opacity=0.00,] (193.51, 53.83) --
	(233.44, 53.83);

\draw[color=drawColor,line width= 1.8pt,line cap=rect,line join=round,fill opacity=0.00,] (202.50, 53.83) --
	(215.02, 53.83);

\draw[color=drawColor,line width= 1.2pt,line cap=rect,line join=round,fill opacity=0.00,] (208.16, 55.89) --
	(208.16, 51.77);
\definecolor[named]{drawColor}{rgb}{0.50,0.50,0.50}

\draw[color=drawColor,line width= 0.6pt,line cap=rect,line join=round,fill opacity=0.00,] (195.50, 48.34) --
	(241.87, 48.34);

\draw[color=drawColor,line width= 1.8pt,line cap=rect,line join=round,fill opacity=0.00,] (205.41, 48.34) --
	(220.08, 48.34);

\draw[color=drawColor,line width= 1.2pt,line cap=rect,line join=round,fill opacity=0.00,] (211.99, 50.40) --
	(211.99, 46.28);
\definecolor[named]{drawColor}{rgb}{0.00,0.00,0.00}

\node[color=drawColor,anchor=base east,inner sep=0pt, outer sep=0pt, scale=  1.00] at ( 23.58, 18.37) {$\sigma_{p}$};

\draw[color=drawColor,line width= 0.6pt,line cap=rect,line join=round,fill opacity=0.00,] (215.83, 26.37) --
	(245.48, 26.37);

\draw[color=drawColor,line width= 1.8pt,line cap=rect,line join=round,fill opacity=0.00,] (224.36, 26.37) --
	(234.42, 26.37);

\draw[color=drawColor,line width= 1.2pt,line cap=rect,line join=round,fill opacity=0.00,] (229.17, 28.43) --
	(229.17, 24.31);
\definecolor[named]{drawColor}{rgb}{0.50,0.50,0.50}

\draw[color=drawColor,line width= 0.6pt,line cap=rect,line join=round,fill opacity=0.00,] (225.70, 20.87) --
	(259.47, 20.87);

\draw[color=drawColor,line width= 1.8pt,line cap=rect,line join=round,fill opacity=0.00,] (235.84, 20.87) --
	(247.12, 20.87);

\draw[color=drawColor,line width= 1.2pt,line cap=rect,line join=round,fill opacity=0.00,] (241.27, 22.93) --
	(241.27, 18.81);
\definecolor[named]{drawColor}{rgb}{0.00,0.00,0.00}

\draw[color=drawColor,line width= 0.2pt,dash pattern=on 4pt off 4pt ,line cap=round,line join=round,fill opacity=0.00,] (178.32,  7.14) --
	(178.32,171.94);

\draw[color=drawColor,line width= 0.2pt,line cap=round,line join=round,fill opacity=0.00,] ( 30.95,171.94) --
	(325.70,171.94);

\draw[color=drawColor,line width= 0.2pt,dash pattern=on 4pt off 4pt ,line cap=round,line join=round,fill opacity=0.00,] ( 30.95,117.00) --
	(325.70,117.00);

\draw[color=drawColor,line width= 0.2pt,dash pattern=on 4pt off 4pt ,line cap=round,line join=round,fill opacity=0.00,] ( 30.95, 89.54) --
	(325.70, 89.54);

\draw[color=drawColor,line width= 0.2pt,dash pattern=on 4pt off 4pt ,line cap=round,line join=round,fill opacity=0.00,] ( 30.95, 62.07) --
	(325.70, 62.07);

\draw[color=drawColor,line width= 0.2pt,dash pattern=on 4pt off 4pt ,line cap=round,line join=round,fill opacity=0.00,] ( 30.95, 34.61) --
	(325.70, 34.61);
\definecolor[named]{fillColor}{rgb}{0.90,0.90,0.90}

\draw[fill=fillColor,draw opacity=0.00,] ( 30.95, 89.54) --
	(178.32, 89.54) --
	(178.32,117.00) --
	( 30.95,117.00) --
	( 30.95, 89.54) --
	cycle;

\draw[color=drawColor,line width= 0.2pt,dash pattern=on 4pt off 4pt ,line cap=round,line join=round,fill opacity=0.00,] ( 30.95,117.00) --
	(325.70,117.00);

\draw[fill=fillColor,draw opacity=0.00,] ( 30.95, 34.61) --
	(178.32, 34.61) --
	(178.32, 62.07) --
	( 30.95, 62.07) --
	( 30.95, 34.61) --
	cycle;

\draw[color=drawColor,line width= 0.2pt,dash pattern=on 4pt off 4pt ,line cap=round,line join=round,fill opacity=0.00,] ( 30.95, 89.54) --
	(325.70, 89.54);

\draw[fill=fillColor,draw opacity=0.00,] ( 30.95,  7.14) --
	(178.32,  7.14) --
	(178.32, 34.61) --
	( 30.95, 34.61) --
	( 30.95,  7.14) --
	cycle;

\draw[color=drawColor,line width= 0.2pt,dash pattern=on 4pt off 4pt ,line cap=round,line join=round,fill opacity=0.00,] ( 30.95, 62.07) --
	(325.70, 62.07);

\node[color=drawColor,anchor=base,inner sep=0pt, outer sep=0pt, scale=  1.00] at ( 30.95,176.97) {-2};

\node[color=drawColor,anchor=base,inner sep=0pt, outer sep=0pt, scale=  1.00] at (104.64,176.97) {-1};

\node[color=drawColor,anchor=base,inner sep=0pt, outer sep=0pt, scale=  1.00] at (178.32,176.97) {0};

\node[color=drawColor,anchor=base,inner sep=0pt, outer sep=0pt, scale=  1.00] at (252.01,176.97) {1};

\node[color=drawColor,anchor=base,inner sep=0pt, outer sep=0pt, scale=  1.00] at (325.70,176.97) {2};

\draw[color=drawColor,line width= 0.2pt,line cap=round,line join=round,fill opacity=0.00,] ( 30.95,171.94) --
	( 30.95,174.68);

\draw[color=drawColor,line width= 0.2pt,line cap=round,line join=round,fill opacity=0.00,] (104.64,171.94) --
	(104.64,174.68);

\draw[color=drawColor,line width= 0.2pt,line cap=round,line join=round,fill opacity=0.00,] (178.32,171.94) --
	(178.32,174.68);

\draw[color=drawColor,line width= 0.2pt,line cap=round,line join=round,fill opacity=0.00,] (252.01,171.94) --
	(252.01,174.68);

\draw[color=drawColor,line width= 0.2pt,line cap=round,line join=round,fill opacity=0.00,] (325.70,171.94) --
	(325.70,174.68);
\end{scope}
\end{tikzpicture}